\documentclass[%
 reprint,
 amsmath,amssymb,
 aps,
prc,
]{revtex4-2}

\usepackage{color}
\usepackage{graphicx}
\usepackage{dcolumn}
\usepackage{bm}
\usepackage{hyperref}
\usepackage[mathlines]{lineno}
\usepackage{xspace}
\usepackage{mathrsfs}

\newcommand{\pT}{$p_T$\xspace}

\newcommand{\kT}{$k_T$\xspace}
\newcommand{\deltapT}{$\delta p_{T}$\xspace}

\newcommand{\GeV}{GeV/$c$\xspace}

\newcommand{\vn}{$v_{n}$\xspace}
\newcommand{\vnum}[1]{$v_{#1}$\xspace}

\newcommand{\sref}[1]{Sec.~\ref{#1}}
\newcommand{\fref}[1]{Fig.~\ref{#1}}
\newcommand{\tref}[1]{Tab.~\ref{#1}}
\newcommand{\eref}[1]{Eq.~\ref{#1}}

\newcommand{\Fref}[1]{Figure~\ref{#1}}

\newcommand{\Eref}[1]{Equation~\ref{#1}}

\newcommand{\pp}{$p$+$p$\xspace}

\newcommand{\Au}{Au+Au\xspace}

\newcommand{\Pb}{Pb+Pb\xspace}

\newcommand{\sNN}{$\sqrt{s_{NN}}$\xspace}

\newcommand{\pikp}{$\pi^{\pm}$, $K^{\pm}$, $p$ and $\bar{p}$\xspace}

\definecolor{UTOrange}{rgb}{1, 0.51, 0.0}
\newcommand{\BG}{{TennGen}\xspace}

\begin{document}


\title{Model studies of fluctuations in the background for jets in heavy ion collisions}

\author{Charles Hughes} 
\author{Antonio Carlos Oliveira da Silva} 
\author{Christine Nattrass} 
\affiliation{University of Tennessee, Knoxville, TN, USA-37996.}


\begin{abstract}
Jets produced in high energy heavy ion collisions are quenched by the quark gluon plasma.  Measurements of these jets are influenced by the methods used to suppress and subtract the large, fluctuating background and the assumptions inherent in these methods.  We compare the measurements of the background in \Pb collisions at \sNN = 2.76 TeV by the ALICE Collaboration \underline{(B. Abelev \textit{et al.}, J. High Energy Phys. $\mathbf{2012}$, 053)} to calculations in \BG (a data-driven random background generator) and PYTHIA Angantyr.  A detailed understanding of the width of these fluctuations is important for reducing uncertainties due to unfolding and extending measurements to lower momenta and larger resolution parameters.
The standard deviation of the energy in random cones in \BG is approximately in agreement with the form predicted in the ALICE paper, with deviations of 1--6\%.
The standard deviation of energy in random cones in Angantyr exceeds the same predictions by approximately 13\%.  Deviations in both models can be explained by the assumption that the single-particle $d^2N/dy$ $dp_T$ is a gamma distribution in the derivation of the prediction, whereas the model uses a different distribution.  This indicates that model comparisons are potentially sensitive to the treatment of the background. We demonstrate that unfolding methods used to remove background fluctuations from jets can affect the comparisons between models and data, $\emph{even in the absence of detector effects}$. Our findings suggest the need to more carefully consider methods for comparing simulations and data. 

\end{abstract}

\maketitle

\section{Introduction}

A hot, dense, strongly interacting liquid of quarks and gluons called the Quark Gluon Plasma (QGP) is briefly created in high energy heavy ion collisions~\cite{Adcox:2004mh,Adams:2005dq,Back:2004je,Arsene:2004fa}.  Two of the key signatures of the formation of the QGP are hydrodynamical flow and jet quenching. The strong azimuthal asymmetry in the final state particles' momenta is a signature of hydrodynamical flow. There are many measurements of jets which can, in principle, provide quantitative constraints on the properties of the medium~\cite{Connors:2017ptx}.  While there have been some constraints on the properties of the medium from measurements of jets~\cite{Burke:2013yra,JETSCAPE:2021ehl}, the era of quantitative measurements is just beginning.  

Improving quantitative constraints on the medium using jet measurements requires a quantitative understanding of the background.  The correlations due to flow lead to an anisotropic background, which can in turn influence jet measurements.  At the Relativistic Heavy Ion Collider (RHIC), mixed events were able to successfully describe the background in measurements of hadron-jet correlations~\cite{Adamczyk:2017yhe}, indicating that the background is dominated by random combinations of particles.  Studies of the background at the Large Hadron Collider (LHC) by the ALICE Collaboration found that the distribution of background energy density measured by using random cones with the leading jet removed were described well by predictions for a random background with correlations due to flow~\cite{ALICE:2012nbx}.

We study the measurements in~\cite{ALICE:2012nbx} in two models. We compare to a data-driven random background generator, \BG \cite{TennGenGITHUB}, which uses the measured single-particle spectra and flow to generate a realistic background without any jets. We also use PYTHIA Angantyr~\cite{Bierlich:2018xfw}, a Monte Carlo generator based on PYTHIA 8.2\cite{Bierlich:2018xfw,Sjostrand:2006za}, which models heavy ion collisions as a superposition of nucleon-nucleon collisions. 
We stress that an understanding of this background is important for reducing uncertainties in jet measurements, which would help extend measurements in heavy ion collisions to higher resolution parameters and lower momenta.  The uncertainties due to unfolding are driven by the width of the distribution rather than the overall level of the background.

We emphasize that while models may simplify the physics of heavy ion collisions they still contain background and background fluctuations. We examine different approaches to unfolding to correct for background fluctuations in models.
We discuss how the presence of this background can affect observables in Monte Carlo simulations, underscoring the need for a treatment of background in model studies similar to that in data. 

\section{Simulations}
\subsection{\BG}

The measured single particle double differential spectra for \pikp from~\cite{Abelev:2012hxa} are fit to a Boltzmann-Gibbs blast wave distribution~\cite{Ristea:2013ara,Schnedermann:1993ws}
\begin{equation}
\begin{split}
\dfrac{d^{2}N}{dp_{T}dy} = Np_{T}\int_{0}^{1} r^{\prime{}}dr^{\prime{}}\left(\sqrt{m^{2} + p_{T}^{2}}\right)\\
\times I_{0}\left(\dfrac{p_{T}\sinh\left(\tanh^{-1}\left(\beta_{s}{r}^{\prime{}^{n}}\right)\right)}{T_{kin.}}\right)\\
\times K_{1}\left(\dfrac{\sqrt{m^{2} + p_{T}^{2}}\cosh\left(\tanh^{-1}\left(\beta_{s}{r}^{\prime{}^{n}}\right)\right)}{T_{kin.}}\right),
\end{split}
\label{eq:BlastWave}
\end{equation}
where \pT is the transverse momentum, $y$ is the rapidity, $N$ is the normalization, $m$ is the mass of the particle, $\beta_{s}$ is the surface velocity, $n$ is an exponent describing the evolution of the velocity profile, and $T_{kin.}$ is the kinetic freeze-out temperature.  The $I_{0}$ and $K_{1}$ are modified Bessel functions.  The reduced radius, $r^{\prime{}}$, is integrated over from 0 to 1. 
The multiplicity of each particle species is determined from charged particle ratios~\cite{Abelev:2013vea} and is scaled up assuming a constant charged particle multiplicity per unit pseudorapidity, $dN_{ch}/d\eta$.  This is a reasonable approximation for the pseudorapidity region used in this analysis, $-0.9 < \eta < 0.9$.  The multiplicities are determined from measurements of the charged particle multiplicities in ALICE at the LHC~\cite{Aamodt:1313050}.  Only the centrality bins in~\cite{Abelev:2013vea} are available (0--5\%, 5--10\%, 10--20\%, 20--30\%, 30--40\%, and 40--50\%) and there are no fluctuations in the multiplicity within a centrality bin.  Only charged hadrons are generated for this analysis. Furthermore, all particles produced from \BG are uncorrelated except through correlations with the event planes.

The azimuthal asymmetry in heavy ion collisions is decomposed using
\begin{equation}
\dfrac{dN}{d\phi} = \dfrac{N_{0}}{2\pi}\left( 1 + \sum_{n=1}^{5}2v_{n}\cos(n(\phi - \Psi_{n})) \right),
\label{eq:Vn}
\end{equation}
where $N_0$ is the number of particles, the \vn coefficients are defined as \vn = $\bigl \langle \cos{ \left[ n \left( \phi - \Psi_{n} \right) \right]} \bigr \rangle$ , and $\phi$ is the azimuthal position of the track.  
The symmetry planes $\Psi_{n}$ are set to zero for even $n$ for simplicity. This differs from the physical correlations between the second and fourth event plane, which have been observed to fluctuate relative to each other~\cite{Aad:2014fla}. While this difference between the \BG simulation and measurements would affect observables sensitive to flow, the simulation is only intended to capture most of the correlations due to flow and not intended as an exact quantitative reproduction. The $\Psi_{n}$ for odd $n$ are randomly thrown from a flat distribution for the odd $n$, roughly matching correlations observed in data~\cite{Aad:2014fla}.  A random \pT is thrown from the distribution in  \eref{eq:BlastWave}, which is then used to determine the \vn.  This is used to construct an azimuthal distribution of particles at the momentum \pT and a random $\phi$ is drawn from that distribution. This is repeated for all the particles in the event. The \vn can also be set to zero to remove the impact of correlations due to flow, leaving a uniform distribution of particles.

When the \vn are included, the \pT-dependent \vn from~\cite{Adam:2016nfo} are fit to a polynomial for $n>1$.  For $n=1$, a rapidity-even \vnum{1} comparable to \vnum{2} and \vnum{3} has been observed~\cite{Luzum:2010fb,ATLAS:2012at,Retinskaya:2012ky}, but it is difficult to measure and is still poorly constrained.  To roughly match these measurements, we use $v_1(p_T) = v_2 (p_T) -0.02$, which will give a negative \vnum{1} for low \pT and a positive \vnum{1} for high \pT, roughly conserving momentum.   The azimuthal coordinate is then randomly drawn from \eref{eq:Vn}.  
The pseudorapidity ($\eta$) is randomly drawn from a uniform distribution for $\mid\eta\mid<$ 0.9. 
For each centrality bin and combination of \vn, 60000 events are generated.  For the 0--10\% centrality bin, the 0--5\% and 5--10\% bins are combined. The code for \BG is available on Github~\cite{TennGenGITHUB}.

\subsection{Angantyr}

PYTHIA Angantyr~\cite{Bierlich:2018xfw} is a Monte Carlo model for heavy ion collisions included in PYTHIA 8~\cite{Bierlich:2018xfw,Sjostrand:2006za}.  It is primarily a superposition of nucleon-nucleon collisions and includes
inelastic collisions, single-diffractive, double-diffractive, and absorptive collisions using a model with fluctuating radii.  The fluctuating nucleon radii result in a fluctuating nucleon-nucleon cross section. This further results in multiplicity fluctuations. Angantyr includes hard scatterings, event-by-event multiplicity fluctuations, and multiparton interactions. Angantyr does not contain flow (string shoving is not enabled in this analysis) or jet quenching.  As such, it is a good baseline for collisions in the absence of a QGP.

Default parameters are used and $20\times10^{3}$ minimum bias \Pb collisions at \sNN = 2.76 TeV and $20\times10^{3}$ minimum bias \Au collisions at \sNN = 200 GeV were generated.
The centrality is determined using the centrality class implemented in Rivet~\cite{Buckley:2010ar}, which uses the multiplicity in the forward pseudorapidity regions matching the ALICE V0-A (2.8 $<$ $\eta$ $<$ 5.1) and V0-C (-3.7 $<$ $\eta$ $<$ -1.7) acceptance \cite{ALICE:2004ftm} and bins the events in terms of the multiplicity in these regions in Angantyr.  
\subsection{Reconstruction efficiency}\label{subsec:reco_eff}
The measurements in~\cite{ALICE:2012nbx} did not include corrections for detector effects so we implement an approximate single track reconstruction efficiency, the dominant effect, to make these model calculations more realistic.  We use a parametrized \pT-dependent efficiency roughly matching the efficiency of the ALICE detector in ~\cite{Abelev:2014ffa} when comparing to~\cite{ALICE:2012nbx}.

\section{Results}

\subsection{Background density $\rho$}
\begin{table}
\caption{
FastJet parameters used}
\begin{ruledtabular}
\begin{tabular}{cccccccc}
\hline
$R_{param}$& 0.4 \\
ghost max. rapidity&  2.0 \\
repeat& 1  \\
ghost area& 0.005  \\
grid scatter& 1.0  \\
$p_{T}$ scatter& 0.1  \\
$<p_{T}^{ghost}>$& $10^{-100}$ GeV/$c$  \\
\end{tabular}
\end{ruledtabular}
\label{table:FJparam}
\end{table}

To match the analysis in~\cite{ALICE:2012nbx}, the background density $\rho$ is estimated using the \kT jet finding algorithm implemented in FastJet~\cite{Cacciari_2012} with the $p_{T}$ recombination scheme and a resolution parameter of $R=0.4$. Reconstructed charged particles with \pT $>$ 0.15 \GeV are input into the jet finder and ghost particles are used to estimate the jet area, $A$.  Jet finding parameters are summarized in \tref{table:FJparam}.  For jet candidates with $|\eta|<0.5$, the median $p_{T}^{jet}/A$ is used to estimate the background momentum density $\rho$ for each event (as in \cite{Cacciari:2007fd}). For Angantyr, the two leading jet candidates are excluded from the sample when calculating the median, as done in~\cite{ALICE:2012nbx}.  Leading jets are not excluded in \BG because it contains no hard scattering.  We simulate the impact of the single track reconstruction efficiency in these calculations.
 
\begin{figure}
    \centering
    \includegraphics[width=1.0\columnwidth]{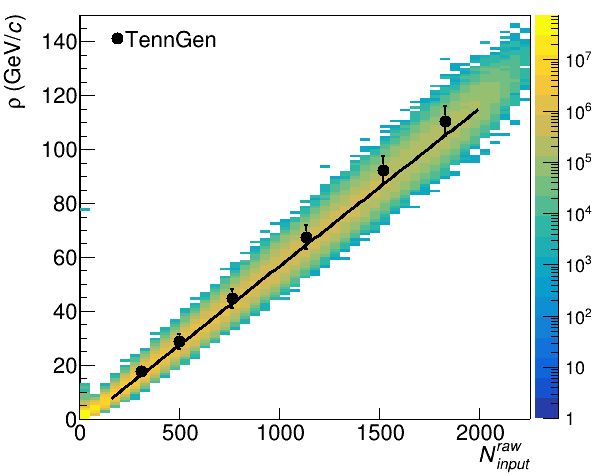}
    \caption{Median event-by-event $\rho$ vs. $N_{input}^{raw}$ for \BG (black points) and Angantyr (z-axis) \Pb collisions at \sNN = 2.76 TeV.  The line is from the fit of a straight line to Angantyr. Parameters from fits to a straight line are in \tref{table:rho_vs_Nraw_fit_params}. 
    }
    \label{fig:Rho_vs_Nraw}
\end{figure}

 \Fref{fig:Rho_vs_Nraw} shows $\rho$ versus the reconstructed number of tracks $N_{input}^{raw}$ for \BG and Angantyr.  These are fit to a straight line with the parameters given in \tref{table:rho_vs_Nraw_fit_params} and compared to fits from~\cite{ALICE:2012nbx}.  The multiplicity dependence is comparable to ALICE data in both models.  Note that the data cover a wider range of multiplicities because \BG only includes fixed multiplicities and Angantyr underestimates the multiplicity distribution by 5--10\%~\cite{Bierlich:2018xfw}.  This difference in the multiplicity means that neither model is directly comparable to the data.  We therefore emphasize comparisons to expectations for a random background in the following sections.
 
 \begin{table}
\caption{
$\rho$ vs multiplicity fit parameters}
\begin{ruledtabular}
\begin{tabular}{cccccccc}
\hline
&Slope& Intercept \\
Angantyr&0.0585 $\pm$ 0.0002&-1.67 $\pm$ 0.09 \\
\BG&0.0610 $\pm$ 0.0029&-1.31 $\pm$ 2.38  \\
ALICE data \cite{ALICE:2012nbx}&0.0623 $\pm$ 0.0002&-3.3 $\pm$ 0.3\\
\end{tabular}
\end{ruledtabular}
\label{table:rho_vs_Nraw_fit_params}
\end{table}

\subsection{Distribution of \deltapT}\label{Sec:deltapTsec}
The soft background does not make jet measurements difficult because it is large, but because it fluctuates, which leads to large, jet-by-jet fluctuations.  This smears the reconstructed jet energy.  This smearing is corrected for in data and, because the background and its fluctuations are present in models which simulate the entire event, it also must be corrected for to make valid comparisons to Monte Carlo models which simulate the entire event.  We therefore investigate the distribution of these background fluctuations and compare them to ALICE measurements of background fluctuations.

Two random cones with a radius $R=0.4$ are drawn within $|\eta|<$ 0.5 for each event.  The \pT of all reconstructed charged hadrons in the cone are added and the background density $\rho$ estimated from jets found with the \kT jet finder is subtracted to get
\begin{equation}
    \delta p_{T} = p_{T,cone} - A_{cone} \rho,
\label{eq:DeltaPt}
\end{equation}
where $A_{cone} = \pi R^2$.
The distribution of \deltapT is a measure of the fluctuations in the background.

\begin{figure}
    \centering
    \includegraphics[width=1.0\columnwidth]{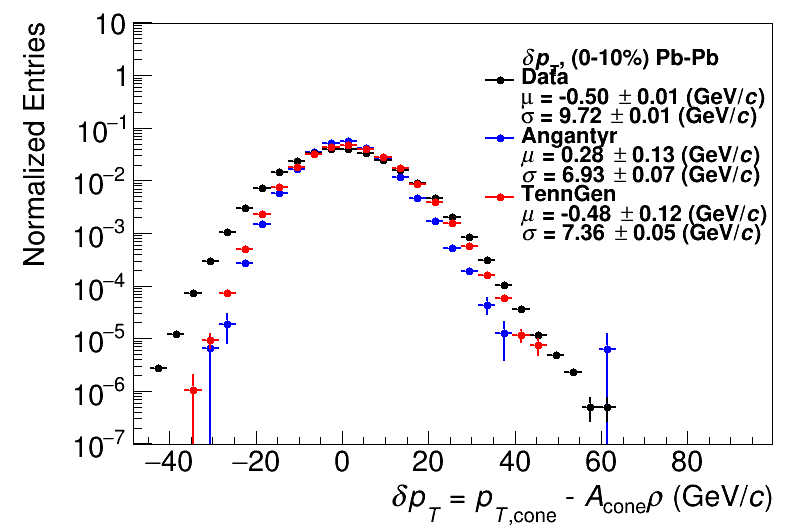}
    \caption{Comparison of \BG and Angantyr to 0--10\% central \Pb collisions at \sNN = 2.76 TeV data from ~\cite{ALICE:2012nbx}. In both Angantyr and the ALICE data, the leading jet has been excluded.}
    \label{fig:Angantyr_BF_Data}
\end{figure}

\Fref{fig:Angantyr_BF_Data} shows the distribution of \deltapT in 3 different sets of points: ALICE data in \Pb collisions at \sNN = 2.76 TeV~\cite{ALICE:2012nbx}, \BG, and Angantyr. The leading jet has been excluded from both the ALICE data and Angantyr. Even though \BG uses ALICE single-particle spectra and \vn, the distributions do not overlap.  This is in part because \BG uses the average multiplicity and does not include fluctuations in the number of particles, leading to a somewhat narrower distribution than the data.  Furthermore, \BG contains no hard processes, resonances, mini-jets or decays. Angantyr contains multiplicity fluctuations and the previously mentioned processes but underestimates the event multiplicity. 

\subsection{Width of the \deltapT distribution}
Understanding the width of fluctuations in the background is important for possible improvements in methods, since the width of the background fluctuations drives the uncertainties from unfolding.
In~\cite{ALICE:2012nbx}, the width of the \deltapT distribution is compared to predictions assuming only fluctuations in the number of particles in the random cone and their momenta and correlations in the distribution of background particles due to flow.  There were small deviations between data and the predictions, but it is not possible to isolate the source of these deviations with studies of data alone.  We compare our model to the same predictions.

The distribution of the sum of momenta from a random sample of particles is discussed in~\cite{TANNENBAUM200129}, where it is applied to distributions of transverse energy in events.  These derivations were applied in~\cite{ALICE:2012nbx} to the problem of random cones.  In~\cite{TANNENBAUM200129}, the single-particle $p_{T}$ spectrum is approximated as a gamma distribution
\begin{equation}
    \frac{d^2N}{dydp_T} \propto \frac{k}{\Gamma(p)}(k p_T)^{p-1} e^{-k p_T},
\end{equation}
where $p \approx 2$ and $k$ are constants and $\Gamma(p)=p!$ if $p$ is an integer.  The $N$-fold convolution of this distribution is itself another gamma distribution with a mean given by $N \langle p_{T} \rangle$ and standard deviation $\sqrt{N} \sigma_{p_T}$.  The \deltapT distribution in \fref{fig:Angantyr_BF_Data} can therefore be fit to a gamma distribution to extract the width.

When there are Poissonian fluctuations in the number of particles in the sample, the distribution is a sum of gamma distributions, with a standard deviation given by
\begin{equation}
    \sigma_{\delta p_{T}} = \sqrt{N \sigma^{2}_{p_{T}} + N  \langle p_{T} \rangle^{2}}.
\label{eq:DeltaPtwidths_random}
\end{equation}
For both \BG with \vn=0 and Angantyr, the distribution of the number of particles in the random cone were consistent with a Poissonian distribution. Appendix~\ref{Sec:AppendixTannenbaum} includes a detailed derivation of \eref{eq:DeltaPtwidths_random} and Appendix~\ref{Sec:ArbitraryDistribution} investigates how \eref{eq:DeltaPtwidths_random} would change if the \pT spectrum were more complicated than a single gamma distribution.

The presence of hydrodynamic flow in~\eref{eq:Vn} leads to non-Poissonian number fluctuations.  If the fluctuations from each term are approximated as uncorrelated and constant as a function of momentum, the width is given by
\begin{equation}
    \sigma_{\delta p_{T}} = \sqrt{N \sigma^{2}_{p_{T}} + (N + 2 N^2 \sum_{n=1}^{\infty} v_n^2) \langle p_{T} \rangle^{2}}.
\label{eq:DeltaPtwidths_flow}
\end{equation}
\noindent In~\cite{ALICE:2012nbx}, only $n=2$ and $n=3$ terms were considered.  These assumptions could be sources of deviations between \eref{eq:DeltaPtwidths_flow} and the observed widths. In addition \eref{eq:DeltaPtwidths_flow} assumes that the $v_{n}$ terms are independent of \pT. For the calculations of \eref{eq:DeltaPtwidths_flow} compared to \BG in this analysis, the un-weighted average $v_{n}$ from \BG is used.
Note that $N$ in \eref{eq:DeltaPtwidths_random} and \eref{eq:DeltaPtwidths_flow} is the number of particles in the random cone, not the charged particle multiplicity in the event.  In Appendix ~\ref{subsec:azimuth_anisotropy} we investigate the impact of flow in greater detail.

\subsubsection{\BG}

\begin{figure}
    \centering
    \includegraphics[width=0.9\columnwidth]{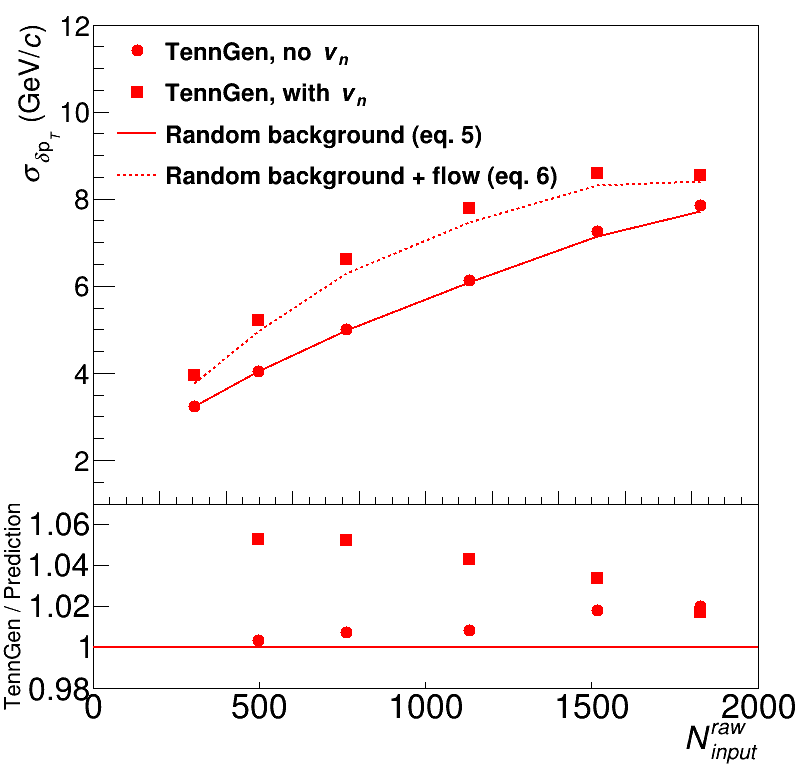}
    \caption{Comparison of the $\delta p_{T}$ distribution's width in \BG with \vn=0 compared to \eref{eq:DeltaPtwidths_random} and nonzero \vn compared to \eref{eq:DeltaPtwidths_flow}. \BG is generated from fits to single particle $p_{T}$ spectra measured in \sNN = 2.76 TeV \Pb collisions with ALICE.}
    \label{fig:BkgGenSim_over_eq}
\end{figure}

\Fref{fig:BkgGenSim_over_eq} shows $\sigma_{\delta p_{T}}$ in \BG with \vn=0 compared to \eref{eq:DeltaPtwidths_random} and with non-zero \vn compared to \eref{eq:DeltaPtwidths_flow}.  The predictions from \eref{eq:DeltaPtwidths_random} and \eref{eq:DeltaPtwidths_flow} use $N$, $\langle p_{T} \rangle$, and $\sigma_{p_{T}}$ in \BG.  The slight deviations seen here are qualitatively consistent with~\cite{ALICE:2012nbx}, but the absence of any correlations other than flow makes the discrepancy easier to interpret in \BG.  The derivation of \eref{eq:DeltaPtwidths_random} assumed that the single particle spectra were a gamma distribution while \BG uses a blast wave, which could explain the roughly 2\% deviation between \BG with \vn=0 and \eref{eq:DeltaPtwidths_random}.  This indicates that the width is dependent on the shape of the spectrum.  The derivation of \eref{eq:DeltaPtwidths_flow} assumed that both the \vn are independent of \pT and that there are no correlations between number fluctuations due to flow, explaining the deviations as high as 6\% between this prediction and \BG with nonzero flow. The derivations in Appendices \ref{Sec:AppendixTannenbaum} and \ref{Sec:ArbitraryDistribution} confirm that these effects can explain the deviations from \eref{eq:DeltaPtwidths_random} and \eref{eq:DeltaPtwidths_flow}.

\subsubsection{Angantyr}

\begin{figure}
    \centering
    \includegraphics[width=\columnwidth]{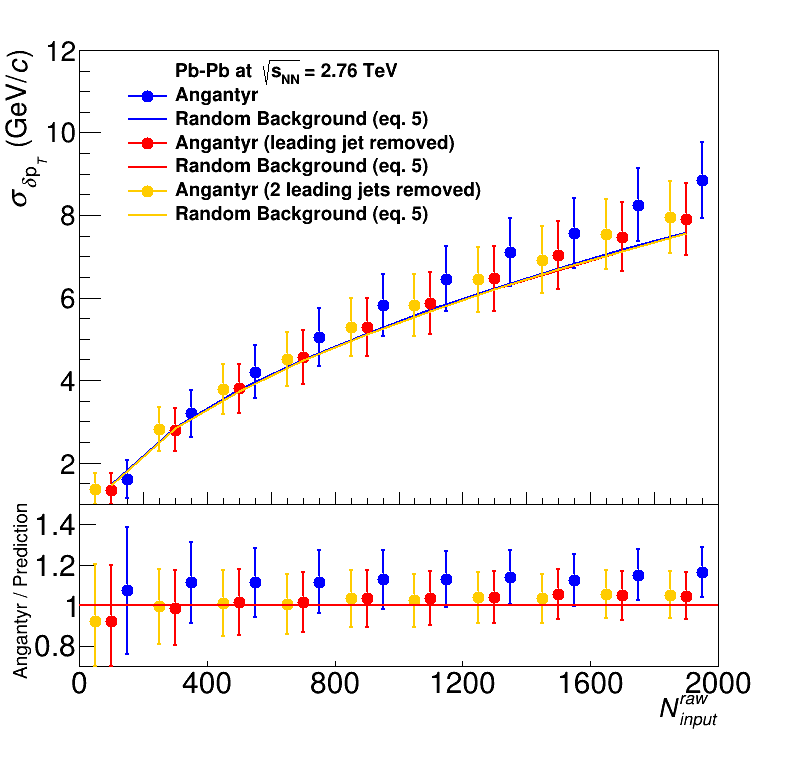}
    \caption{Comparison of the $\delta p_{T}$ distribution's width in Angantyr for \Pb collisions at \sNN = 2.76 TeV with \eref{eq:DeltaPtwidths_random} with zero, one, and two leading jets omitted from the sample.}
    \label{fig:Angantyr_LeadingJets}
\end{figure}

\Fref{fig:Angantyr_LeadingJets} compares the $\delta p_{T}$ widths in Angantyr with no jets excluded, the leading jet excluded, and the leading two jets excluded from the sample to  \eref{eq:DeltaPtwidths_random}. Leading jets are excluded by requiring a large separation between the axis of the random cone and the anti-$k_{T}$ jet axis, $\Delta R = \sqrt{(\phi_{jet}-\phi_{cone})^2+(\eta_{jet}-\eta_{cone})^2}>1.0$.  The predictions from \eref{eq:DeltaPtwidths_random} use the $N$, $\langle p_{T} \rangle$, and $\sigma_{p_{T}}$ in Angantyr. The widths in Angantyr have an average difference of 12\% with respect to what is predicted by \eref{eq:DeltaPtwidths_random} when no leading jets are removed. The discrepancy gets smaller when jets are excluded from the sample. The average differences are 4\% and 3\% when one or two leading jets are removed, respectively.

\begin{figure}
    \centering
        \includegraphics[width=0.9\columnwidth]{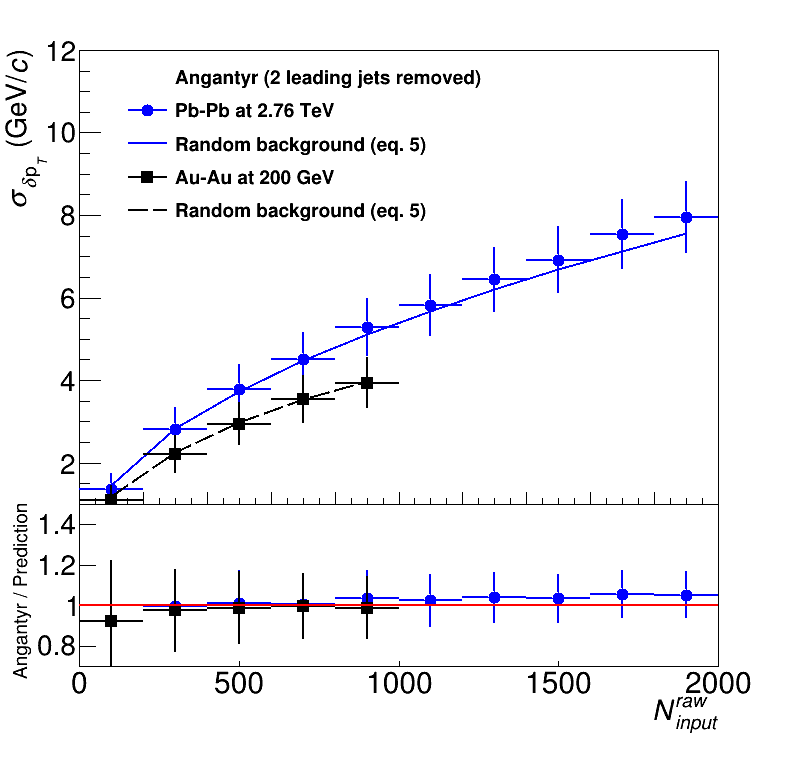}
    \caption{Comparison of the $\delta p_{T}$ distribution's width in Angantyr (two leading jets removed from the sample) for \Pb collisions at \sNN = 2.76 TeV and \Au collisions at \sNN = 200 GeV compared to \eref{eq:DeltaPtwidths_random}.}
    \label{fig:Angantyr_over_eq3}
\end{figure}

\Fref{fig:Angantyr_over_eq3} shows the $\delta p_{T}$ widths in \Pb collisions at \sNN = 2.76 TeV and \Au collisions at \sNN = 200 GeV with the two leading jets removed from the sample. The predictions from \eref{eq:DeltaPtwidths_random} use the $N$, $\langle p_{T} \rangle$, and $\sigma_{p_{T}}$ from Angantyr at each energy. The average difference from the prediction in \Au is 2\%. The lower energy should have fewer jets than the higher energy, which could partially explain why the $\sigma_{\delta p_{T}}$ is closer to the prediction in \Au. Additional differences could be from the difference between the particle $p_{T}$ spectrum in Angantyr and a gamma distribution.

\begin{figure}
    \centering
    \includegraphics[width=0.9\columnwidth]{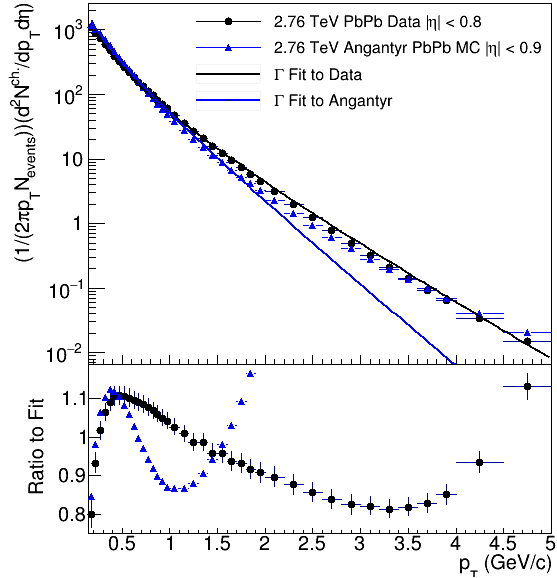}
    \caption{Comparison of gamma distribution fits to 10--20\% central 2.76 TeV Pb+Pb data and 10--20 \% central 2.76 TeV Angantyr Monte Carlo. 
    } 
    \label{fig:Angantyr_Data_Gamma_Fit}
\end{figure}

\Fref{fig:Angantyr_Data_Gamma_Fit} shows fits of the particle spectra from \Pb collisions at \sNN = 2.76 TeV  in data and in Angantyr to a gamma distribution.  Since these are single particle spectra, it is not possible to remove particles from jets.  The gamma distribution describes the data better than it describes Angantyr.   This indicates that the deviations between Angantyr and predictions from \eref{eq:DeltaPtwidths_random} shown in \Fref{fig:Angantyr_LeadingJets} may be largely due to the difference in the shapes of the spectra, as supported by the calculations in Appendix~\ref{Sec:ArbitraryDistribution}.


\subsection{Unfolding jets in heavy-ion collision Monte Carlo}\label{sec:unfolding}

Experiments correct for the migration of jets from their correct momentum bin to another bin, which distorts the momentum, using a procedure called unfolding~\cite{DAgostini:2010hil, Hocker:1995kb}.   In experiment, this smearing arises due to both detector effects and jet background fluctuations. Usually, a response matrix is determined from full simulation of the detector response to a jet.  A simulated PYTHIA \pp collision may be embedded in data from a heavy ion collision in order to use a data-driven smearing due to fluctuations in the background; we construct a similar response matrix using a \Pb collision from Angantyr and call this the ``embedding response matrix".  
Alternatively, a response matrix including both effects can be determined by multiplying two response matrices, one describing the detector response to a jet in \pp collision and one describing background fluctuations observed in the data~\cite{ALICE:2015mjv}.  In the absence of detector effects, as is the case in these studies, the only effect is from background fluctuations; we construct a similar response matrix for Angantyr using the fluctuations in \fref{fig:Angantyr_BF_Data} and call this the ``fluctuation-only response matrix".  
We also construct a response matrix using a PYTHIA \pp event embedded in an Angantyr \Pb event but only using the particles from the \pp event to determine the measured jet momentum, and then multiply this matrix by the background fluctuation response matrix.  This should capture changes in the behavior of the jet finder in a heavy ion collision while still maintaining the assumption that the impact of changes in the behavior of the jet finder and fluctuations in the background can be factorized.  We call this the ``fluctuation plus matching response matrix."  




Two sets of jets are reconstructed, one containing only charged particles from the \pp event, which is considered the generated distribution, and another using all charged particles in either event, which is considered the smeared distribution. In order to build the response matrix, a match between the jets in the two sets is established by requiring  a bijective match and $\Delta R = \sqrt{\Delta\eta^{2} + \Delta\phi^{2}} < 0.45$ where $\Delta\eta$ and $\Delta\phi$ are the differences in $\eta$ and $\phi$ between the generated and smeared jets.  We do not include the impact of the finite single track reconstruction efficiency.

\begin{figure*}
    \centering
    \resizebox{\textwidth}{!}{
    \includegraphics[width=0.9\columnwidth]{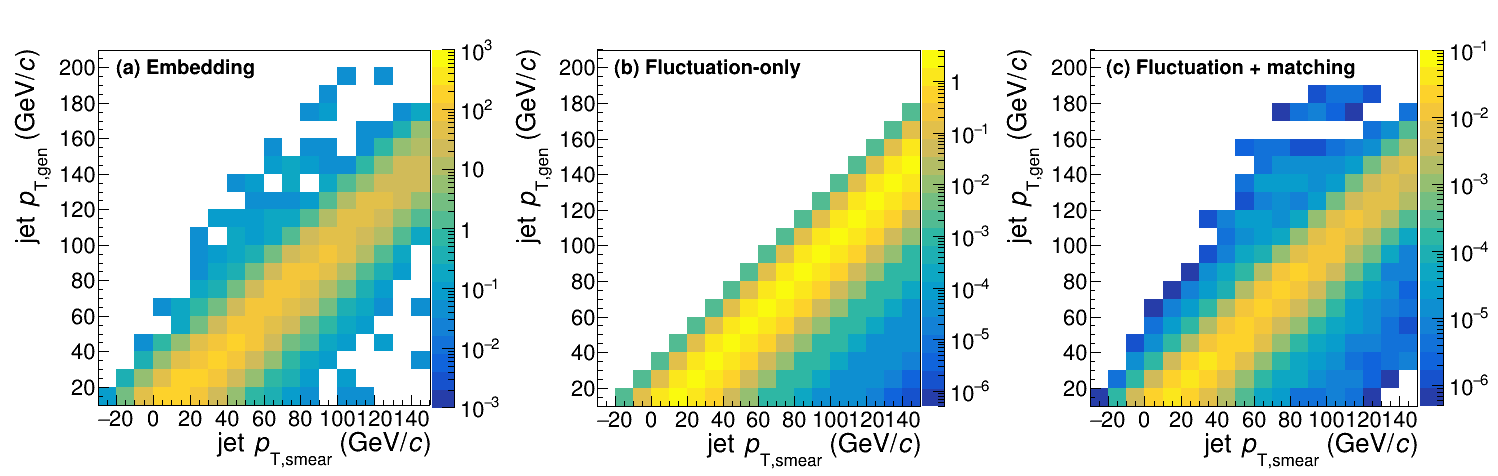}
    }
    \caption{(a) Embedding, (b) fluctuation-only, and (c) fluctuation plus matching response matrices}
    \label{fig:responsematrices}
\end{figure*}

The response matrices are shown in \fref{fig:responsematrices}.  The fluctuation-only response matrix does not describe jets reconstructed well in the region where the reconstructed momenta is below the true momenta, which can be seen in the embedding response matrix, and it predicts a significant contribution from jets reconstructed well above their true momenta, which is not evident in the embedding response matrix.  The fluctuation plus matching response matrix is also unable to capture this behavior.

In order to demonstrate closure, an unfolded distribution has to converge to the true distribution.   We use only the jets in the combined \Pb plus \pp event which were matched to a jet in the \pp event and unfold this transverse momentum spectrum using the different response matrices.  We compare this to the true transverse momentum spectrum in PYTHIA \pp events.
We use Bayesian unfolding implemented in the RooUnfold~\cite{Adye:2011gm} package. The singular value decomposition method was used as a cross-check and all results were consistent with those obtained with Bayesian unfolding. 



\begin{figure}
    \centering
    \includegraphics[width=1.0\columnwidth]{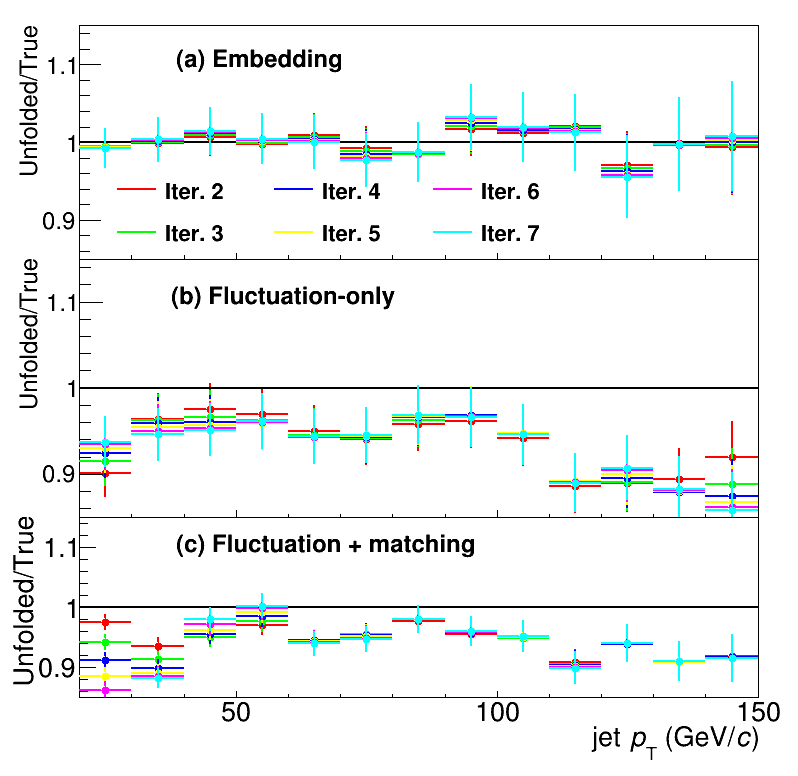}
    \caption{Unfolded spectra over the true distribution of the (a) embedding (b) fluctuation-only and (c) fluctuation plus matching response matrices}
    \label{fig:UnfoldedRatios}
\end{figure}

The ratios of the unfolded spectra to the true distributions are shown in Fig. \ref{fig:UnfoldedRatios}. The number of iterations in the Bayesian unfolding procedures is varied from 2 to 7. The results using the embedding response matrix converge quickly, with little change after the second iteration, and this procedure successfully recovers the true distribution to within around 4\% . The results using the fluctuation-only response matrix change more and the difference between the results and the true distribution is about 5\% below 110 GeV/$c$, increasing to about 10\% above that.  The results from the fluctuation plus matching response matrix are comparable to those with the fluctuation-only response matrix.

The spectra unfolded using the fluctuation-only and the fluctuation plus matching response matrices are systematically lower than the true spectra.  This could skew the interpretation of comparisons between models and data.
This indicates that there is an interplay between background fluctuations and the behavior of the jet finder in a heavy ion environment, and that robust comparisons between data and full Monte Carlo models may require not just unfolding, but a response matrix created using embedding, just like the procedure for data analysis.

\section{Discussion}


Our first observation is that a large, fluctuating background uncorrelated with jet measurements is, indeed, present in Monte Carlos which simulate the full event, necessitating background subtraction in such models. 

Other models, such as JEWEL ~\cite{Zapp:2008gi} nominally give the user only the particles from signal jets with only some ambiguity as to which particles from the medium were influenced by the jet. This ambiguity can be approached by looking at the two extremes, when only particles from the hard parton shower are included and when  medium particles which have interacted with the jet are included~\cite{KunnawalkamElayavalli:2016lzw}.  


That works well if the only possible uncertainty is theoretical. However, the field has made incomplete assumptions about the background for jet measurements in the past, for instance omitting \vnum{3} from the background for dihadron correlations, leading to several erroneous observations~\cite{Connors:2017ptx}.  Biases, or even mistakes, in measurements due to incomplete assumptions about the background subtraction are possible.  We therefore advocate following the philosophy of RIVET~\cite{Bierlich:2020wms}: the exact same strategy should be used in both the analysis and the Monte Carlo model, to the extent possible, in order to ensure that the comparison between data and the model is valid.  In practice, this means that the background subtraction should be implemented in Monte Carlo.  In dihadron correlations, this would have resulted in valid comparisons between models and data, even though the result would be more sensitive to the soft background than was intended in the measurement.  For models such as JEWEL, which nominally contain only or mostly particles directly from the jet signal, aspects of the background subtraction could still be applied, particularly when they might impose a bias in the measurement.  For example, reflection about $\eta=0$ for the background subtraction ~\cite{Chatrchyan:2014ava,Chatrchyan:2012gw} could be implemented in JEWEL.

For Monte Carlo models which simulate the full event, such as Angantyr, this means that the exact same background subtraction method should be applied to the model as is used in data, including the corrections for fluctuations.  Background fluctuations are significantly different in models and in data, so it is not sufficient for comparisons to use experimental observations of the fluctuations or an experimental response matrix.  This poses some complications for uncorrected measurements, even if the response matrix is provided with the measurement, as the fluctuations will be different in the model and the data.  The sensitivity of the shape of the background to subtle differences in the shape of the spectrum and the details of correlations between particles in background poses particular problems for comparisons between data and models.  It implies that the corrections for the background fluctuations must be done separately for each model, using a method consistent with that used in data. Moreover, we find that it is necessary to unfold using a response matrix constructed in the same way as in the measurement.



\section{Conclusions}
While our studies broadly support the conclusions in~\cite{ALICE:2012nbx} that the background fluctuations are dominated by random combinations of particles, we find that this width is sensitive to both details of the hydrodynamical background and the shape of the single particle momentum spectrum.  These effects are less than 6\% for \BG, a data-driven random background generator, and around 13\% in PYTHIA Angantyr depending on how many jets are removed from the events.


As measurements of jets in heavy-ion collisions reach higher precision, it is important to make sure that models are comparable to data.  Some of the details of flow correlations would be difficult to fully describe in background subtraction methods.  Area-based subtraction techniques such as those used by ALICE with a data-driven determination of the fluctuations~\cite{Abelev:2013kqa,ALICE:2015mjv} and the $\eta$-reflection method used by CMS~\cite{Chatrchyan:2014ava,Chatrchyan:2012gw} should be robust to these effects.  

It is less clear how these subtle effects in the width of fluctuations in the background would be incorporated into mixed events~\cite{Adamczyk:2017yhe} or impacted by the iterative subtraction techniques used by CMS~\cite{Khachatryan:2016jfl} and ATLAS~\cite{Aad:2012vca}.  Many models, such as Angantyr, may not accurately reproduce the background in heavy ion collisions.  Implementation of the full experimental method in model calculations, using tools such as Rivet, is essential for robust and meaningful comparisons between models and data.  

We note that there are models which do not attempt to simulate the full event, such as JEWEL~\cite{Zapp:2008gi}, as well as models where it is possible to separate jet ``signal."  This may reduce the direct sensitivity to background models, but it adds an additional theoretical uncertainty, since arbitrary distinctions between ``signal" and ``background" must be made.  This may be less problematic for certain observables, particularly those less sensitive to soft radiation; however, it is precisely those observables which are likely to be most interesting for studies of partonic energy loss.  We therefore urge care in comparisons between data and models, reproducing as many parts of the experimental method as possible. 


\section{Acknowledgements}
We are grateful to Christian Klein-B\"osing, Marco van Leeuwen, Leif L\"onnblad, Mike Tannenbaum, and Gary Westfall for productive discussions and Alexandre Shabetai and Ejiro Umaka for feedback on the manuscript.  This work was supported in part by funding from the Division of Nuclear Physics of the U.S. Department of Energy under Grant No. DE-FG02-96ER40982 and from the National Science Foundation under Grant No. OAC-1550300.  We also acknowledge support from the UTK and ORNL Joint Institute for Computational Sciences Advanced Computing Facility.


\bibliographystyle{unsrt}
\bibliography{Bibliography}

\appendix*
 {
\onecolumngrid
\section{Derivations}
\subsection{Mean and width of the sum of particles drawn from a $\Gamma$ distribution}\label{Sec:AppendixTannenbaum}
The derivation of the distribution of energies in~\cite{TANNENBAUM200129} and the associated widths in~\cite{ALICE:2012nbx} assumed a spectrum of the form
\begin{equation}\label{Eq:Tannenbaum}
    \frac{dN}{dp_T} = a p_T^b e^{-c p_T},
\end{equation}
where $N$ is the number of particles, \pT is the  transverse momentum, and $a$, $b$, and $c$ are constants.  The normalized probability distribution, $dP^1/dp_T$, for the probability of a single random particle drawn from the distribution having a momentum \pT takes the same form, with $a = \frac{c^{b+1}}{\Gamma (b+1) }$. As in ~\cite{TANNENBAUM200129}, it is convenient to parametrize \ref{Eq:Tannenbaum} in terms of the mean (b = $\dfrac{\mu^2}{\sigma^2}$) and the variance (c = $\dfrac{\mu}{\sigma^2}$):
\begin{equation}\label{Eq:TannenbaumMeanSigma}
    \frac{dN}{dp_T}\left(\mu , \sigma, p_T\right) = \dfrac{\left( \frac{\mu}{\sigma^2} \right)^{\frac{\mu^2}{\sigma^2}}}{\Gamma (\frac{\mu}{\sigma^2})} p_T^{\frac{\mu^2}{\sigma^2}-1} e^{-\frac{\mu}{\sigma^2} p_T}.
\end{equation}
It can be shown that the parameters $\mu$ and $\sigma$ are in fact the mean and standard of deviation of \ref{Eq:TannenbaumMeanSigma}:
\begin{eqnarray}\label{Eq:MeanofTannenbaumMeanSigma}
    \mu \{ \frac{dN}{dp_T} \}  =  \int_{0}^{\infty} p_T \cdot \frac{dN}{dp_T}\left(\mu , \sigma, p_T\right) \,dp_T = \int_{0}^{\infty} \dfrac{\left( \frac{\mu}{\sigma^2} \right)^{\frac{\mu^2}{\sigma^2}}}{\Gamma (\frac{\mu}{\sigma^2})} p_T^{\frac{\mu^2}{\sigma^2}} e^{-\frac{\mu}{\sigma^2} p_T} \,dp_T = \mu,\\
    \sigma \{ \frac{dN}{dp_T} \}  =  \sqrt{ \int_{0}^{\infty} p_T^2 \cdot \frac{dN}{dp_T}\left(\mu , \sigma, p_T\right) \,dp_T - \left( \mu \{ \frac{dN}{dp_T} \} \right)^2 } = \sqrt{ \sigma^2 + \mu^2 - \mu^2 } = \sigma, \label{Eq:SigmaofTannenbaumMeanSigma}
\end{eqnarray}
The distribution of the sum of the momenta of two particles is given by the convolution where $\frac{dP^1}{dp_T}$ is  \ref{Eq:TannenbaumMeanSigma}:
\begin{equation}
    \frac{dP^2}{dp_{T,tot}} = \int_0^{p_{T,tot}} \frac{dP^1(x)}{dp_T }. \frac{dP^1(p_{T,tot}-x)}{dx }.
\end{equation}
This is repeated each time an additional particle is added, where each iteration is given by
\begin{equation}
    \frac{dP^n}{dp_{T,tot}} = \int_0^{p_{T,tot}} \frac{dP^1(x)}{dp_T } \frac{dP^{n-1}(p_{T,tot}-x)}{dx },
\end{equation}
\noindent where $n$ is the number of particles.
The distribution of the total \pT  in the sample, $p_{T, tot}$,  for $n$ particles drawn from this distribution is given by
\begin{equation}\label{Eq:GammaSumpTinCone}
    \frac{dP^n}{dp_{T,tot}} = \dfrac{\mu^{\frac{n\mu^2}{\sigma^2}}}{\sigma^2 \Gamma \left( \frac{n \mu^2}{\sigma^2} \right)}\left( \dfrac{p_{T,tot}}{\sigma^2} \right)^{\frac{n\mu^2}{\sigma^2} - 1}e^{-\frac{p_{T,tot}\mu}{\sigma^2}},
\end{equation}
with a corresponding mean in total cone $p_{T}$ given by
\begin{equation}\label{Eq:MeanGammaSumpTinCone}
    \mu \{ \frac{dP^n}{dp_{T,tot}} \} = \int_{0}^{\infty} p_{T,tot} \frac{dP^n}{dp_{T,tot}} \,dp_{T,tot} = n \mu,
\end{equation}
and variance 
\begin{equation}\label{Eq:WidthGammaSumpTinCone}
    \sigma^2 \{ \frac{dP^n}{dp_{T,tot}} \} = \int_{0}^{\infty} p_{T,tot}^2 \frac{dP^n}{dp_{T,tot}} \,dp_{T,tot} - \left( \mu \{ \frac{dP^n}{dp_{T,tot}} \} \right)^2  = n \sigma^2 .
\end{equation}
The total width of fluctuations of the sum of $p_T$ in a random cone is the quadrature sum of the  Poissonian fluctuations in the number of particles in the random cone and the width of the n-fold convolution \ref{Eq:WidthGammaSumpTinCone}
\begin{equation}\label{Eq:ALICEWidth}
    \sigma^2 \left( \delta p_T \right) = \left( \sqrt{n} \mu \{ \dfrac{dN}{dp_T} \} \right)^2 + \left( \sigma^2 \{ \frac{dP^n}{dp_{T,tot}} \} \right) = n \mu + n \sigma^2,
\end{equation}
where $n$ is the number of particles in the cone, $\mu$ is the mean $p_T$ of the single-particle distribution, and $\sigma$ is the standard deviation of the $p_T$ of the single-particle distribution. \Eref{Eq:ALICEWidth} is the same as Equation 3 in \cite{Abelev:2012hxa}.

\subsection{Deviations from a $\Gamma$ distribution}\label{Sec:ArbitraryDistribution}
We  note that the $p_T^j e^{-c p_T}$ for integer $j$ form a complete set and we therefore can write an arbitrary spectral shape as 
\begin{equation}\label{Eq:spectraTaylor}
    \frac{dP}{dp_T} =  \sum_{j=0}^{\infty} a_j p_T^j e^{-c p_T}.
\end{equation}
The prescription in Appendix \sref{Sec:AppendixTannenbaum}, in principle, also works for an arbitrary case such as this.
Since a single term is shown in~\cite{ALICE:2012nbx} to work well, we assume a form
\begin{equation}\label{Eq:ApproxProbDist}
    \frac{dN}{dp_T} =   \alpha p_T^a (1+ \beta p_T^b) e^{-c p_T},
\end{equation}
where $\alpha$, $\beta$, $a$, $b$, and $c$ are small and $\beta$ is much smaller than $\alpha$.  Since the measured spectra are approximately a gamma distribution, this should provide a realistic quantification of deviations from a perfect gamma distribution.  Taking into account an explicit normalization [such that the integral of $\ref{Eq:ApproxProbDist}$ over all $p_{T}$ is 1], the mean of $\ref{Eq:ApproxProbDist}$ is given by
\begin{equation}\label{Eq:ApproxProbDistMean}
    \mu \{ \dfrac{dN}{d p_T} \} = \frac{c^b \Gamma (a+2) + \beta \Gamma (a + b + 2)}{c^{1+b}\Gamma (a+1) + c \beta \Gamma (a+b+1)}
\end{equation}
and the variance is given by
\begin{equation}\label{Eq:ApproxProbDistVariance}
    \sigma^{2} \{ \dfrac{dN}{d p_T} \} = \frac{c^b \Gamma (a+3) + \beta \Gamma (a + b + 3)}{c^2(c^b \Gamma (a+1) + \beta \Gamma (a+b+1))} - (\frac{c^b \Gamma (a+2) + \beta \Gamma (a + b + 2)}{c^{1+b}\Gamma (a+1) + c \beta \Gamma (a+b+1)})^2.
\end{equation}
Following the procedure laid out in Appendix \ref{Sec:AppendixTannenbaum}, we find the $n$-fold convolution of \ref{Eq:ApproxProbDist}. We do this using Laplace transforms,  $f^{(n)}(x) = \mathscr{L}^{-1} \{ \mathscr{L} \{ f(x) \}^n \}$, and induction. The first convolution gives
\begin{equation}\label{Eq:ApproxProbDist1stconvolution}
        \begin{split}
        \frac{dP^2}{dp_{T,tot}} = \mathscr{L}^{-1} \{ \mathscr{L} \{ \frac{dN}{dp_T} \}^2 \} = \dfrac{c^{2a + 2b + 2} p_{T,tot}^{2a+1} e^{-c p_{T,tot}}}{(c^b \Gamma (a + 1) + \beta \Gamma (a + b + 1) )^2} \\
        \times\left( \dfrac{ \Gamma^{2} (a+1) }{ \Gamma (2a + 2) } + \dfrac{ 2 p_{T,tot}^b \beta \Gamma (a+1) \Gamma (a+ b + 1) }{ \Gamma (2a + b + 2) } + \dfrac{  p_{T,tot}^{2b} \beta^2 \Gamma^2 (a+ b + 1) }{ \Gamma (2a + 2b + 2)} \right).
        \end{split}
\end{equation}
The 2nd convolution gives:
\begin{equation}\label{Eq:ApproxProbDist2ndconvolution}
    \begin{split}
       \frac{dP^3}{dp_{T,tot}} = \mathscr{L}^{-1} \{ \mathscr{L} \{ \frac{dN}{dp_T} \}^3 \} = \dfrac{c^{3a + 3b + 3} p_{T,tot}^{3a+2} e^{-c p_{T,tot}}}{[c^b \Gamma (a + 1) + \beta \Gamma (a + b + 1) ]^3}\\        
       \times( \dfrac{ \Gamma^{3} (a+1) }{ \Gamma (3a + 3) } + \dfrac{ 3 p_{T,tot}^b \beta \Gamma^2 (a+1) \Gamma (a+ b + 1) }{ \Gamma (3a + b + 3) } \\
        + \dfrac{ 3 p_{T,tot}^{2b} \beta^2 \Gamma (a+1) \Gamma^2 (a+ b + 1) }{ \Gamma (3a + 2b + 3) } + \dfrac{  p_{T,tot}^{3b} \beta^3 \Gamma^3 (a+ b + 1) }{ \Gamma (3a + 3b + 3)} ).
    \end{split}
\end{equation}
By induction the n$th$ convolution is
\begin{equation}\label{Eq:ApproxProbDistnthconvolution}
    \frac{dP^n}{dp_{T,tot}} = \dfrac{c^{n(a+b+1)}p_{T,tot}^{na + n - 1}e^{-cp_{T,tot}}}{(c^b \Gamma(a+1) + \beta \Gamma (a + b + 1))^n} \sum_{m=0}^{n} \dfrac{\binom{n}{m} p_{T,tot}^{bm}\beta^{m} \Gamma^{n-m} (a + 1 ) \Gamma^{m} (a + b + 1)}{\Gamma (na + mb + n )}.
\end{equation}
The variance of \ref{Eq:ApproxProbDistnthconvolution} is
\begin{equation}\label{Eq:VarianceApproxProbDistnthconvolution}
    \begin{split}
    \sigma^2 \{\frac{dP^n}{dp_T,tot} \} = \\  \dfrac{c^{n(a+b+1)}}{(c^b \Gamma(a+1) + \beta \Gamma (a + b + 1))^n} \sum_{m=0}^{n} \dfrac{\binom{n}{m} c^{-na -bm -n - 2}  \beta^{m} \Gamma^{n-m} (a + 1 ) \Gamma^{m} (a + b + 1) \Gamma (na + mb + n + 2)}{\Gamma (na + mb + n )}  \\
    - \dfrac{c^{2n(a+b+1)}}{(c^b \Gamma(a+1) + \beta \Gamma (a + b + 1))^{2n}} \left( \sum_{m=0}^{n} \dfrac{\binom{n}{m} c^{-na -bm -n - 1}  \beta^{m} \Gamma^{n-m} (a + 1 ) \Gamma^{m} (a + b + 1) \Gamma (na + mb + n + 1)}{\Gamma (na + mb + n )} \right)^{2}. \\
    \end{split}
\end{equation}
When combined with the Poissonian fluctuations in the number of particles in the cone, this can gives the width of the fluctuations for the sum of momentum in the cone as a function of $n$ particles in the cone:
\begin{equation}\label{Eq:WidthSumpTinconeApproxProbDist}
    \begin{split}
    \sigma^2 \left( \delta p_T \right)  = \left( \sqrt{n} \mu \{ \dfrac{dN}{d p_T} \} \right)^2 +  \sigma^2 \{\frac{dP^n}{dp_{T,tot}} \}  = \\
    \left( \sqrt{n} \frac{c^b \Gamma (a+2) + \beta \Gamma (a + b + 2)}{c^{1+b}\Gamma (a+1) + c \beta \Gamma (a+b+1)} \right)^2 + \\
    \dfrac{c^{n(a+b+1)}}{(c^b \Gamma(a+1) + \beta \Gamma (a + b + 1))^n} \sum_{m=0}^{n} \dfrac{\binom{n}{m} c^{-na -bm -n - 2}  \beta^{m} \Gamma^{n-m} (a + 1 ) \Gamma^{m} (a + b + 1) \Gamma (na + mb + n + 2)}{\Gamma (na + mb + n )}  \\
    - \dfrac{c^{2n(a+b+1)}}{(c^b \Gamma(a+1) + \beta \Gamma (a + b + 1))^{2n}} \left( \sum_{m=0}^{n} \dfrac{\binom{n}{m} c^{-na -bm -n - 1}  \beta^{m} \Gamma^{n-m} (a + 1 ) \Gamma^{m} (a + b + 1) \Gamma (na + mb + n + 1)}{\Gamma (na + mb + n )} \right)^{2}. \\
    \end{split}
\end{equation}

This expression is difficult to simplify so that the impact of deviations from a Gamma distribution can be interpreted easily.  Instead we use realistic numbers and show the impact in~\fref{fig:Additional_Term_Gamma}.  Deviations from a single gamma distribution always increase the width, with the deviations increasing monotonically from that of a single gamma distribution.  Thus we see that realistic deviations from a gamma distribution increase the width of the distribution of momenta in random cones.

\begin{figure}
    \centering
    \includegraphics[width=1.0\columnwidth]{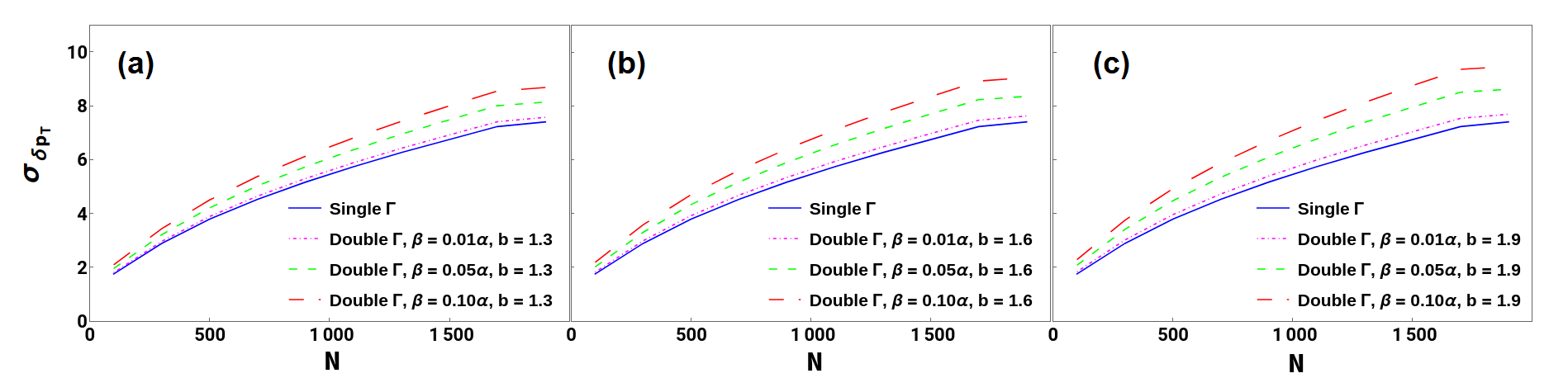}
    \caption{Comparison of the widths of background fluctuations for particles drawn from a single $\Gamma$ distribution in $p_{T}$ from Eq. \ref{Eq:Tannenbaum} with realistic parameters to widths of background fluctuations (Eq. \ref{Eq:WidthSumpTinconeApproxProbDist}) derived from double $\Gamma$ distributions in $p_{T}$ with the same parameters for $a$, $\alpha$, and $c$ as Eq. \ref{Eq:Tannenbaum} but varying $\beta$ and $b$ parameters. The single $\Gamma$ distribution is the same function in each panel, $N$ is the multiplicity of the collision, and $\sigma p_{T}$ is the width of the jet background fluctuations. }
    \label{fig:Additional_Term_Gamma}
\end{figure}

In addition, we constructed gamma distributions with the same mean and standard deviation as the spectra in Angantyr and \BG for \Pb events at \sNN = 2.76 TeV.  We drew several samples of the average number of particles observed in random cones for each generator and added up the total momentum. The track momentum distributions and the distribution of the total momenta are given in \fref{fig:DistributionThrowing} and the properties of these distributions are given in \tref{table:rho_vs_Nraw_fit_params2}.  This exercise isolates the impact of the shape of the spectra alone.  The shifts in the mean of the sums of all momenta are small.  The shift in the standard deviation of the sum of all momenta from the true distribution to the gamma distribution is small for both, but larger for Angantyr.  This also demonstrates that the shapes of the spectra are important for describing fluctuations in the background.

\begin{table}
\caption{Mean ($\mu_{\Sigma p_{T}}$) and standard deviation ($\sigma_{\Sigma p_{T}}$) for distribution of total momenta of the average number of particles in a random cone ($N_A$) for 10--20\% central \BG \Pb events at \sNN = 2.76 TeV and \Pb events at \sNN = 2.76 TeV with Angantyr with a multiplicity of 1200--1400, as well as for gamma distributions with the same mean momenta and standard deviations.}
\begin{ruledtabular}
\begin{tabular}{cccccccc}
\hline
&$N_{A}$  &$\mu_{\Sigma p_{T}}$ (\GeV) & $\sigma_{\Sigma p_{T}}$ (\GeV)\\
Angantyr&58&37.54 $\pm$ 0.02&6.29 $\pm$ 0.01 \\
Angantyr~$\Gamma$&58&36.12 $\pm$ 0.02&6.12 $\pm$ 0.01 \\
TennGen&52&35.68 $\pm$ 0.02&6.12 $\pm$ 0.01   \\
TennGen~$\Gamma$ &52&35.70 $\pm$ 0.01&6.11 $\pm$ 0.01\\
\end{tabular}
\end{ruledtabular}
\label{table:rho_vs_Nraw_fit_params2}
\end{table}

\begin{figure*}[h]
\centering
    \includegraphics[width=0.47\linewidth]{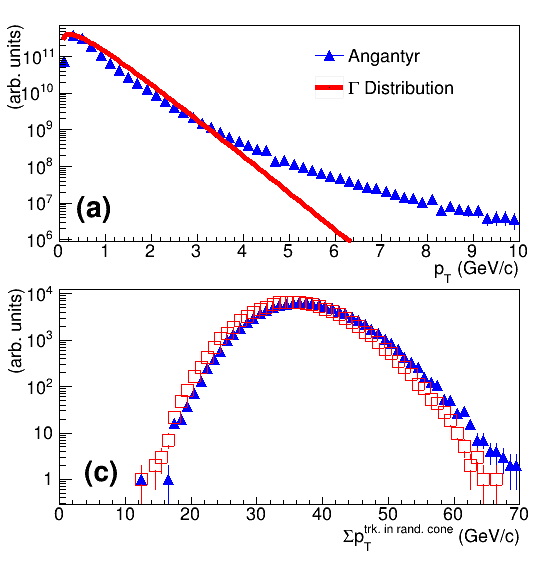}
    \includegraphics[width=0.47\linewidth]{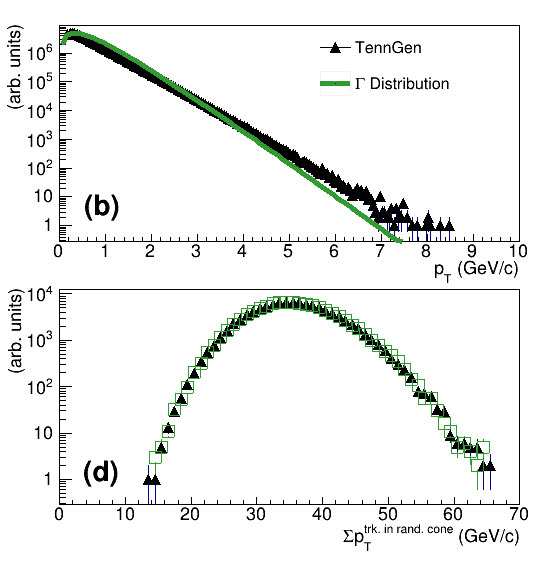}
\caption{Charged track momentum distributions $\frac{dN}{dp_T}$ for (a) Angantyr events with a multiplicity of 1200--1400 and (b) 10--20\% central \BG events  for \Pb collisions at \sNN=2.76 TeV and Gamma Distributions with the same means and variances. Distributions of the sum of track momenta for $N_{A}=58$ for Angantyr and $N_{A}=35.7$ for \BG for both the $\frac{dN}{dp_T}$ in the model and the gamma distribution with the same single track mean and standard deviation.}\label{fig:DistributionThrowing}
\end{figure*}

\subsection{Including azimuthal anisotropy}\label{subsec:azimuth_anisotropy}
We consider the azimuthal anisotropy due to \vn and show that this is a special case of that derived in \sref{Sec:ArbitraryDistribution}.  The standard expression of azimuthal anisotropy in a heavy ion collision is
\begin{equation}\label{Eq:vnexpansionsimple}
    \frac{d^2P}{dp_Td(\phi-\psi_n)} = A\left(1 + 2 \sum_{n=1}^{\infty} v_n(p_T) \cos[n (\phi-\psi_n)]\right),
\end{equation}
where $A$ is a normalization factor, $\phi$ is the azimuthal angle of a particle's momentum vector, $\psi_n$ is the azimuthal position of the $n$th order event plane, and the $v_n$ are the $n$th order azimuthal anisotropies.  Without loss of generality, we can express the momentum dependence of the \vn with a Taylor expansion
\begin{equation}\label{Eq:vnTaylor}
    v_n = \sum_{m=0}^{\infty} v_{n,m} p_T^m ,
\end{equation}
where the $v_{n,m}$ are constants so that \eref{Eq:vnexpansionsimple} can be rewritten as
\begin{equation}\label{Eq:vnexpansioncomplex}
    \frac{d^2N}{dp_Td(\phi-\psi_n)} = \frac{dN}{dp_T} (1 + 2 \sum_{n=0}^{\infty} v_{n,0} \cos(n (\phi-\psi_n)) + 2 p_T \sum_{n=0}^{\infty}  v_{n,1} \cos(n (\phi-\psi_n)) + ...).
\end{equation}
The dominant $v_{n,m}$ can be chosen so that this can be rewritten in the form of \eref{Eq:spectraTaylor}.  If only the first term is kept, corresponding to constant $v_n$, the momentum and azimuthal dependencies factorize.  The analysis in Appendix~\ref{Sec:AppendixTannenbaum} can be applied to the momentum dependence.  The mean is given by
\begin{eqnarray}
    \mu \{ \frac{dN}{dp_T} \}  = \Big(\frac{1}{2\pi} \int_0^{2\pi} \Big(1 + 2 \sum_{n=0}^{\infty} v_{n,0} \cos(n (\phi-\psi_n))\Big)   d\phi \Big) \Big( \int_{0}^{\infty} p_T \frac{dN}{dp_T} \,dp_T \Big) =\mu.
\end{eqnarray}
The average does not change because the average over all azimuthal angles is one.
The standard deviation  is given by
\begin{eqnarray}
    \sigma \{ \frac{dN}{dp_T} \}  =  \sqrt{\Big( \frac{1}{2\pi} \int_0^{2\pi} \Big(1 + 2 \sum_{n=0}^{\infty} v_{n,0} \cos(n (\phi-\psi_n))\Big) d\phi \Big)  \int_{0}^{\infty} p_T^2 \frac{dN}{dp_T}  \,dp_T  - \mu^2 } = \sigma .
\end{eqnarray}
This also does not change.  The change in standard deviation due to $v_n$ in \eref{eq:DeltaPtwidths_flow} is entirely because of the change in the number of particles.  However, including a single momentum dependent term in the $v_n$ in \eref{Eq:vnexpansioncomplex} increases deviations of $\frac{dN}{dp_T}$ from a single gamma distribution, increasing the width.  Since the $v_n$ are momentum-dependent, any realistic $v_n$ will increase the width.
}

\end{document}